\documentclass[prc,superscriptaddress,unsortedaddress,twocolumn]{revtex4}
\usepackage{graphicx}
\usepackage{amsmath}
\usepackage{amssymb}
\usepackage{times}
\usepackage{bm}
\usepackage{color}
\usepackage{ulem}

\def\ga{\mathrel{\mathpalette\fun >}}
\def\fun#1#2{\lower3.6pt\vbox{\baselineskip0pt\lineskip.9pt
\ialign{$\mathsurround=0pt#1\hfil##\hfil$\crcr#2\crcr\sim\crcr}}}
\newcommand{\beq}{\begin{equation}}
\newcommand{\eeq}{\end{equation}}
\newcommand{\bea}{\begin{eqnarray}}
\newcommand{\eea}{\end{eqnarray}}

\def\ve{\varepsilon}

\def\non{\nonumber}
\def\dfrac#1#2{{\displaystyle\frac{#1}{#2}}}

\begin{document}

\title{Quantum mechanical description of excitation energy
distribution of the reaction residue
\\ in nucleon-induced inclusive one-nucleon knockout reactions
}

\author{Kazuyuki Ogata}
\email[]{kazuyuki@rcnp.osaka-u.ac.jp}
\affiliation{Research Center for Nuclear Physics (RCNP), Osaka
University, Ibaraki 567-0047, Japan}

\date{\today}

\begin{abstract}
\noindent
{\bf Background:}
Understanding of inclusive one-nucleon knockout reactions for
long-lived fission fragments (LLFPs) is crucial for nuclear
transmutation studies. However, the particle and heavy ion
transport code system (PHITS) severely overshoots the inclusive
one-nucleon knockout cross sections $\sigma_{-1N}$. \\
{\bf Purpose:}
Development of a reaction model for describing the inclusive
one-nucleon knockout processes is necessary. A key is
specification of the position and the momentum of a nucleon
inside a nucleus to be struck by the incident nucleon. \\
{\bf Methods:}
The semiclassical distorted wave model incorporating
the Wigner transform of the one-body nuclear density matrix
is applied to the calculation of excitation energy distributions
of reaction residues. Decay of a residue is described by
introducing a threshold parameter for the minimum excitation energy
of it. \\
{\bf Results:}
With reasonable values of the parameter, the measured
$\sigma_{-1N}$ for several LLFPs are reproduced by the
proposed reaction model.
The incident energy dependence of $\sigma_{-1N}$ is found to
be governed by that of the nucleon-nucleon cross sections at
energies higher than about 75~MeV. At low energies,
the nuclear absorption and the Coulomb penetrability
also become important.
The energy dependence of neutron-induced $\sigma_{-1N}$
is predicted and found
to be quite different from that of proton induced
one. \\
{\bf Conclusions:}
The proposed reaction model is shown to be promising in
discussing the energy dependence of nucleon-induced
inclusive one-nucleon knockout processes. The energy dependence
of the measured $\sigma_{-1p}$ for $^{107}$Pd above 100~MeV is,
however, not explained by the present calculation.
\end{abstract}

\maketitle

\section{Introduction}
\label{sec1}

Reduction of high-level radioactive wastes produced in nuclear power
plants is one of the most crucial issues in modern society.
As a possible solution, so far a great effort
has been devoted to realize the nuclear transmutation
technology~\cite{IAEA04}.
Very recently, nuclear spallation cross sections
for some long-lived fission products (LLFPs),
which are very important reaction data for nuclear transmutation,
have been measured for the first time at the RIKEN RI Beam Factory
(RIBF)~\cite{Wan16,Wan17,Kaw17}.
However, it is very difficult to measure the cross section data
for all LLFPs at various incident energies. We therefore need a
reliable model that can describe the existing data and has
a predictive power.
The particle and heavy ion transport code system (PHITS)~\cite{PHITS}
is one of the most promising candidates for {\lq\lq}a standard model''
to evaluate not only cross sections for specific reaction processes
but also the amount of reaction products of a macroscopic system.
In fact, PHITS has successfully been applied to facility design,
medical physics, radiation production, and geoscience.

It was found that in general PHITS reproduces well the
spallation cross sections taken at RIBF with hydrogen and deuterium
targets.
However, for processes in which only one nucleon is removed,
PHITS tends to significantly overshoot the cross sections in many
cases. It was discussed in Ref.~\cite{Man15} that overshooting
of one-nucleon knockout cross sections by
the Li\`ege intranuclear cascade
(INCL) model~\cite{Bou02,Bou13}, which is incorporated in PHITS,
comes from the nucleon momentum distribution inside
a nucleus adopted by the model.
In the INCL model a nucleon in the nuclear surface must have higher
momentum, being in direct contradiction to the picture of the local
Fermi-gas model. 
As a result, after a nucleon-nucleon collision that knocks out
one nucleon, the excitation energy of the reaction residue
has to be too low to allow further particle emission.
The one-nucleon knockout cross sections evaluated with the INCL model
are thus significantly larger than the experimental data.
Very recently, a similar overshooting was reported~\cite{Par17}
for the spallation cross sections of $p$-$^{136}$Xe at 200 MeV/nucleon
taken at GSI. In Ref.~\cite{Man15} a phenomenological prescription
to ease the restriction on nucleon momentum distributions in
the nuclear surface was proposed, which improved the agreement between
the result of the INCL model and experimental data.
Thus, the nucleon momentum distribution inside a nucleus is
found to be a key for evaluating one-nucleon knockout cross sections.

In this paper I propose a quantum-mechanical reaction model for describing
one-nucleon knockout processes, with incorporating the Wigner transform
(WT) of the one-body density matrix (OBDM) as a distribution of positions
and momenta of a nucleon inside a nucleus.
The reaction model is based on the semiclassical distorted wave
(SCDW) model~\cite{LK91,KM92,Wat99,Oga99,Sun99,Oga02,Oga03}
developed for describing inclusive ($p,p'$) and ($p,n$)
processes to the continuum.
Although the SCDW model can include up to three-step processes in
terms of the nucleon-nucleon collision, in this paper I take into account
only a one-step process. This limitation will not significantly affect
the discussion on one-nucleon knockout processes, because these
occur only when a reaction residue has excitation energy in
a narrow window that allows a reaction system to emit just one nucleon.
Multistep direct processes hardly satisfy the condition required,
at least for kinematics with which a meaningfully large cross
section is obtained.

It should be noted that the reaction model proposed in this study
is different from a usual distorted wave impulse approximation
(DWIA)~\cite{CR77,CR83,Wak17}.
The DWIA assumes a single-particle (s.p.) wave function for the
struck nucleon, and evaluates a triple-differential cross section,
momentum distribution of the residual nucleus, or an integrated
cross section, for a nucleon knockout process.
These observables are multiplied by the so-called
spectroscopic factor $S_\mathrm{sp}$ for the s.p. state of interest
and compared with data; in some cases through comparison with data
$S_\mathrm{sp}$ is determined. This model is expected to work when
the experiment is designed to observe a s.p. state (or some specific
s.p. states) of a nucleus. In such measurement, usually the residual
nucleus B is in the ground state or low-lying bound excited states.
On the other hand, in the spallation process what is probed is
no longer pure s.p. states, that is, inclusive measurement of
s.p. structures.
In other words, reaction processes in which B is in continuum states
play a crucial role.
As described below, inclusion of the WT of the OBDM as in Ref.~\cite{Sun99}
allows one to describe the nuclear many-body system in a suitable
manner to calculate continuous excitation energy distribution
of the reaction residue.

This paper is organized as follows. In Sec.~\ref{sec2}
I introduce a reaction model for describing inclusive one-nucleon
knockout processes. In Sec.~\ref{sec31} numerical inputs are given and 
in Sec.~\ref{sec32} I discuss the excitation energy distribution
of reaction residues. The energy dependence of proton-induced
one-nucleon knockout cross sections are shown in Sec.~\ref{sec33}
and those for neutron-induced reactions are discussed in Sec.~\ref{sec34}.
Finally, a summary is given in Sec.~\ref{sec4}.

\section{Formalism}
\label{sec2}

In what follows, unless otherwise denoted, formulation is done in
the center-of-mass (c.m.) frame. For simplicity I adopt nonrelativistic
kinematics; in the actual numerical calculation a relativistic correction
on the kinematics is included.
I consider nucleon ($N$) induced inclusive processes, ($N,N'x$),
where $x$ represents that all the final states except for the outgoing
nucleon are not specified. I assume the projectile nucleon is the same as
the ejectile. This particle is called a leading particle (LP).

In the semiclassical distorted wave (SCDW) model incorporating
the WT of the OBDM, the double
differential cross section (DDX) for ($N,N'x$), with
the outgoing energy $E_f$ and the solid angle $\Omega_f$ specified,
is given by~\cite{Sun99}
\begin{align}
\frac{d^{2}\sigma}{dE_{f}d\Omega_{f}}=
&\; C\int d\boldsymbol{R}\,d\boldsymbol{k}_{\beta}d\boldsymbol{k}_{\alpha}\;
\delta\left(\boldsymbol{K}_{f}+\boldsymbol{k}_{\beta}-\boldsymbol{K}_{i}-\boldsymbol{k}_{\alpha}\right) \non \\
&  \times
\left\vert \chi_{f,\boldsymbol{K}_{f}%
}^{\left(  -\right)  }\left(  \boldsymbol{R}\right)  \right\vert ^{2}\left[
2-f_{\mathrm{h}}^{\left(  \beta\right)  }\left(  \boldsymbol{k}_{\beta},\boldsymbol{R}%
\right)  \right] \non \\
& \times
\left\vert\tilde{t}\left(\boldsymbol{\kappa}^{\prime},\boldsymbol{\kappa}\right)\right\vert^{2}
f_{\mathrm{h}}^{\left(\alpha\right)}\left(\boldsymbol{k}_{\alpha},\boldsymbol{R}\right)
\left\vert \chi_{i,\boldsymbol{K}_{i}}^{\left(  +\right)  }\left(
\boldsymbol{R}\right)  \right\vert ^{2} \non \\
& \times \delta\left(
E_{f}+\varepsilon_{\beta}-E_{i}-\varepsilon_{\alpha}\right),
\label{ddx}
\end{align}%
where
\beq
C=
\frac{\mu_{f}\mu_{i}}{\left(2\pi\hbar^{2}\right)^{2}}
\frac{K_{f}}{K_{i}}\frac{1}{\left(2\pi\right)^{3}}\frac{1}{2}
\eeq
is the kinematical factor; $\mu_c$ ($c=i$ or $f$) is the reduced
mass in the initial ($i$) or final ($f$) channel and
${K}_{c}$ is the asymptotic momentum of the LP in channel $c$.
The distorted wave of the LP in the entrance (exit) channel, which is
solved under an outgoing (incoming) boundary condition, is
denoted by
$\chi_{i,\boldsymbol{K}_{i}}^{(+)}$ ($\chi_{f,\boldsymbol{K}_{f}}^{(-)}$).
$\boldsymbol{k}_{\alpha}$ and $\boldsymbol{k}_{\beta}$ are
what one may interpret as the momenta of the nucleon $N_{\mathrm{t}}$ inside the target
nucleus A (the target nucleon) before and after the collision,
respectively, with the LP.

$\tilde{t}$ represents the matrix element of a nucleon-nucleon
effective interaction calculated with the relative momentum
$\boldsymbol{\kappa}$ ($\boldsymbol{\kappa}^{\prime}$) of the
colliding two nucleons in the initial (final) state;
$\tilde{t}$ is related to the nucleon-nucleon cross section
$(d\sigma / d\Omega)_{NN_{\mathrm{t}}}$ as
\beq
\left\vert \tilde{t}\left(
\boldsymbol{\kappa}^{\prime},\boldsymbol{\kappa}\right)  \right\vert ^{2}%
=
\frac{\left(  2\pi\hbar^{2}\right)  ^{2}}{\mu_{NN_{\mathrm{t}}}^{2}}
\left(
\frac{d\sigma}{d\Omega}
\right)_{NN_{\mathrm{t}}},
\eeq
where $\mu_{NN_{\mathrm{t}}}$ is the reduced mass of the two nucleons.

The initial and final states of the nucleus are denoted by
$\alpha$ and $\beta$, respectively, and these labels are
understood to specify also whether the target nucleon
is proton or neutron.
$f_{\mathrm{h}}^{(\gamma)}$ ($\gamma=\alpha$ or $\beta$) is the WT
of the OBDM for hole states:
\beq
f_{\mathrm{h}}^{\left(\gamma\right)}\left(\boldsymbol{k}_{\gamma},\boldsymbol{R}\right)
\equiv
\sum_{nlj}\frac{2j+1}{2l+1} F_{nlj}
f_{nlj}\left(\boldsymbol{k}_{\gamma},\boldsymbol{R}\right)
\label{wt}
\eeq
with
\begin{align}
f_{nlj}\left(  \boldsymbol{k}_{\gamma},\boldsymbol{R}\right)  \equiv
&\; \int d\boldsymbol{u}\,e^{-i\boldsymbol{k}_{\gamma}\cdot\boldsymbol{u}}
\sum_{m}\varphi_{nlmj}^{\ast}\left(  \boldsymbol{R}-\boldsymbol{u}%
/2\right)
\non \\
& \times \varphi_{nlmj}\left(
\boldsymbol{R}+\boldsymbol{u}/2\right).
\label{pwt}
\end{align}
Here, $\varphi_{nlmj}$ is the spatial part of a single-particle (s.p.) wave
function specified by the principal quantum number $n$, the orbital
angular momentum $l$, its third component $m$, and the total angular
momentum $j$. The summation in Eq.~(\ref{wt}) is taken over all the
occupied states of the target nucleon.
We adopt the filling approximation for an open orbit;
$F_{nlj}$ is the ratio of the number of nucleons in the orbit to
$2j+1$. It is known that WT satisfies the following sum rule
\beq
f_{\mathrm{h}}^{\left(\gamma\right)}\left(\boldsymbol{k}_{\gamma},\boldsymbol{R}\right)
+
f_{\mathrm{p}}^{\left(\gamma\right)}\left(\boldsymbol{k}_{\gamma},\boldsymbol{R}\right)
=2,
\eeq
where $f_{\mathrm{p}}^{(\gamma)}$ is the WT for particle states. This sum rule has been
used in deriving Eq.~(\ref{ddx}) and $2-f_{\mathrm{h}}^{(\beta)}$ appears accordingly.
For more details, readers are referred to Ref.~\cite{Sun99}.

It should be noted that in Eq.~(\ref{ddx}) I have made a further approximation
to the original SCDW model, that is, use of
the asymptotic momentum $\boldsymbol{K}_{c}$ instead of
its local momentum in describing the kinematics of the LP inside A.
The local momentum for the LP~\cite{LK91} is one of the most essential ingredients
of the SCDW model, which allows the LP to reach a classically-inaccessible
region and collide with a target nucleon there. Consequently, DDXs at very
forward and backward angles change significantly as well as those with
large energy transfer. However, in these regions the DDX is much smaller
than its maximum value. A cross section integrated over $E_f$ and $\Omega_f$,
which we are interested in as discussed below, is hardly affected by the
inclusion of the local momentum of the LP.

A noteworthy feature of the SCDW model is that the DDX
is given as an incoherent sum over the coordinate $\boldsymbol{R}$,
that is, the localization of the nucleon-nucleon collision inside A.
This allows one
to define {\it the collision point} of reaction processes,
which gives in part a theoretical foundation to
classical and quantum-mechanical simulations, such as
the INC model~\cite{Bou02,Bou13,Man15,Ser47},
quantum molecular dynamics (QMD)~\cite{Nii95},
and the time-dependent version of antisymmetrized molecular
dynamics (AMD)~\cite{Kan12},
applied to nuclear reaction studies.
With the smoothness approximation to the WT, which was justified in
Ref.~\cite{Sun99}, the s.p. energy $\ve_\gamma$ is evaluated by
\beq
\ve_\gamma=
\frac{\hbar^2}{2m_{N_{\mathrm{t}}}}k_\gamma^2+U_{N_{\mathrm{t}}}(R),
\label{spe}
\eeq
where $m_{N_{\mathrm{t}}}$ is the mass of the target nucleon and
$U_{N_{\mathrm{t}}}$ is the s.p. potential;
for proton $U_{N_{\mathrm{t}}}$ includes the Coulomb interaction.
We then define the following quadruple {\lq\lq}cross section''
\begin{align}
\frac{d^{4}\sigma}{dE_{f}d\Omega_{f}dk_{\alpha}dR}=
&\;
C\frac{k_{\alpha}m_{N_{\mathrm{t}}}}{\hbar^{2}q}
\int d\phi_{\alpha}\int R^{2}d\Omega_{R}\;
\left\vert \chi_{f,\boldsymbol{K}_{f}}^{\left(-\right)}\left(\boldsymbol{R}\right)\right\vert ^{2}
\non \\
&  \times\,
\left[ 2-f_{\mathrm{h}}^{\left(\beta\right)}\left(\boldsymbol{k}_{\beta},\boldsymbol{R}\right)\right]
\left\vert \tilde{t}\left(\boldsymbol{\kappa}^{\prime},\boldsymbol{\kappa}\right)\right\vert ^{2}
\non \\
& \times
f_{\mathrm{h}}^{\left(\alpha\right)}\left(\boldsymbol{k}_{\alpha},\boldsymbol{R}\right)
\left\vert \chi_{i,\boldsymbol{K}_{i}}^{\left(+\right)}\left(\boldsymbol{R}\right)\right\vert ^{2},
\label{qdx}
\end{align}%
where $q$ is the magnitude of the momentum transfer
$\boldsymbol{q}=\boldsymbol{K}_{i}-\boldsymbol{K}_{f}$,
$\Omega_{R}$ is the solid angle of $\boldsymbol{R}$, and
$\phi_{\alpha}$ is the azimuthal angle of
$\boldsymbol{k}_{\alpha}$ with respect to $\boldsymbol{q}$.
Through Eq.~(\ref{spe}), $\ve_\alpha$ is determined by
$R$ and $k_\alpha$. Thus one may construct
\begin{align}
\frac{d^{2}\sigma}{dE_{f}d\ve_{\alpha}}
&\equiv
f_\mathrm{id}
\int d\Omega_{f} dk_{\alpha} dR\;
\frac{d^{4}\sigma}{dE_{f}d\Omega_{f}dk_{\alpha}dR}
\non \\
& \hspace{5mm} \times
\delta
\left(
\frac{\hbar^2}{2m_{N_\mathrm{t}}}k_\gamma^2+U_{N_\mathrm{t}}(R)-\ve_{\alpha}
\right) \non \\
&=
\frac{d^{2}\sigma}{d\omega d\ve_{\alpha}},
\label{ddx2}
\end{align}
where $\omega$ is the energy transfer defined by
\beq
\omega=E_i - E_f=\ve_\beta-\ve_\alpha.
\eeq
In Eq.~(\ref{ddx2}) $f_\mathrm{id}$ is a normalization factor for nucleon-nucleon
cross sections; it is 1/2 (unity) when
the LP is identical to (different from) the target nucleon,

Here I consider two possibilities for the residual nucleus,
which is called the pre-fragment (PF) in PHITS,
to be followed by particle decay.
One is the nucleus $\mathrm{A}-N_{\mathrm{t}}$ referred to as B,
and the other is A.
The first case is realized when
i) $\ve_\beta$ is larger than the barrier height $\ve_{\mathrm{br}N_{\mathrm{t}}}$
for the struck nucleon and
ii) the intrinsic energy of B
\beq
\ve_{\mathrm{ex}}^\mathrm{B} \equiv -S_{N_{\mathrm{t}}}^\mathrm{A}-\ve_\alpha,
\label{eexb}
\eeq
where $S_{N_{\mathrm{t}}}^\mathrm{A}$ is the nucleon separation energy of A,
is larger than a threshold energy $\ve_{0N_{\mathrm{t}}}$.
$\ve_{\mathrm{br}N_{\mathrm{t}}}$ is taken to be the maximum value of $U_{N_{\mathrm{t}}}$;
for neutron $\ve_{\mathrm{br}N_{\mathrm{t}}}=0$.
Since $\ve_{\mathrm{ex}}^\mathrm{B}$ can be
interpreted as excitation energy of B, ideally it should be positive
and $\ve_{0N_{\mathrm{t}}}$ is set to 0. However, in this study I use $\ve_{0N_{\mathrm{t}}}$ as
an adjustable parameter to reproduce one-nucleon knockout cross sections
measured at some specific energies. Necessity of introducing
$\ve_{0N_{\mathrm{t}}}$ will be due to lack of understanding of nuclear structure
wave functions and also to inadequacy of the interpretation of
$\boldsymbol{k}_\gamma$ as a momentum of the target nucleon, that is,
Eq.~(\ref{spe}).

Once B is adopted as the PF, its excitation energy distribution
is obtained by
\beq
\frac{d\sigma^{(\mathrm{PF:B})}}{d\ve_{\mathrm{ex}}^\mathrm{B}}
=
\int d\omega d\ve_{\alpha}\;
\frac{d^{2}\sigma^{(\mathrm{PF:B})}}{d\omega d\ve_{\alpha}}\;
\delta(
 -S_{N_{\mathrm{t}}}^\mathrm{A}-\ve_\alpha
-\ve_{\mathrm{ex}}^\mathrm{B}
),
\label{exdisb}
\eeq
where
\beq
\frac{d^{2}\sigma^{(\mathrm{PF:B})}}{d\omega d\ve_{\alpha}}
\equiv
\frac{d^{2}\sigma}{d\omega d\ve_{\alpha}}
\Theta(\ve_\beta-\ve_{\mathrm{br}N_{\mathrm{t}}})
\Theta(\ve_{\mathrm{ex}}^\mathrm{B}-\ve_{0N_{\mathrm{t}}})
\eeq
with $\Theta$ being the Heaviside function.
Unless the conditions i) and ii) are both satisfied, the nucleus
A is chosen as the PF, and its excitation energy distribution
is given following the definition of $\omega$ by
\beq
\frac{d\sigma^{(\mathrm{PF:A})}}{d\ve_{\mathrm{ex}}^\mathrm{A}}
=
\int d\ve_{\alpha} \;
\frac{d^{2}\sigma^{(\mathrm{PF:A})}}{d\omega d\ve_{\alpha}}
\eeq
with
\begin{align}
\frac{d^{2}\sigma^{(\mathrm{PF:A})}}{d\omega d\ve_{\alpha}}
\equiv & \;
\frac{d^{2}\sigma}{d\omega d\ve_{\alpha}} \non \\
& \times
\left[1-
\Theta(\ve_\beta-\ve_{\mathrm{br}N_{\mathrm{t}}})
\Theta(\ve_{\mathrm{ex}}^\mathrm{B}-\ve_{0N_{\mathrm{t}}})
\right].
\end{align}

Rigorously speaking, the decaying property of the PF should be
described by a statistical model. For example, one can export
$d\sigma^{(\mathrm{PF:C})}/d\ve_{\mathrm{ex}}^\mathrm{C}$ (C is A or B) to
the code system CCONE~\cite{Iwa15} and obtain one-nucleon knockout cross
sections $\sigma_{-1N}$.
Instead of this, I just employ a threshold rule to calculate
$\sigma_{-1N}$ in the following manner.
When the FP is B, I calculate
\beq
\sigma_{-1{N_{\mathrm{t}}}}^{(\mathrm{PF:B})}
\equiv
\int_{\ve_{0N_{\mathrm{t}}}}^{S_\mathrm{min}^\mathrm{B}}
d\ve_{\mathrm{ex}}^\mathrm{B}\;
\frac{d\sigma^{(\mathrm{PF:B})}}{d\ve_{\mathrm{ex}}^\mathrm{B}}
\eeq
with
\beq
S_\mathrm{min}^\mathrm{B}
\equiv
\min (S_{n}^\mathrm{B},S_p^\mathrm{B}).
\eeq
The upper limit $S_\mathrm{min}^\mathrm{B}$
of the integration is set so as not to
occur further particle decay.
On the other hand, when the PF is A,
\beq
\bar{\sigma}_{-1{N_{\mathrm{t}}}}^{(\mathrm{PF:A})}
\equiv
\int_{S_{N_{\mathrm{t}}}^\mathrm{A}}^{S_{N_{\mathrm{t}}}^\mathrm{A}+S_\mathrm{min}^\mathrm{B}}
d\ve_{\mathrm{ex}}^\mathrm{A}\;
\frac{d\sigma^{(\mathrm{PF:A})}}{d\ve_{\mathrm{ex}}^\mathrm{A}}
\eeq
is evaluated. Then for proton I estimate a Coulomb penetration
probability $P_\mathrm{A}^\mathrm{Coul}$ with the WKB approximation, with
averaging out its energy dependence between 0 and
$S_\mathrm{min}^\mathrm{B}$ for the proton energy. The contributions
of the events in which the PF is A to the one-nucleon knockout
cross sections read
\begin{align}
\sigma_{-1p}^{(\mathrm{PF:A})}
&=
\dfrac{P_\mathrm{A}^\mathrm{Coul}}{1+P_\mathrm{A}^\mathrm{Coul}}
\bar{\sigma}_{-1p}^{(\mathrm{PF:A})},
\\
\sigma_{-1n}^{(\mathrm{PF:A})}
&=
\bar{\sigma}_{-1n}^{(\mathrm{PF:A})}
+
\dfrac{1}{1+P_\mathrm{A}^\mathrm{Coul}}
\bar{\sigma}_{-1p}^{(\mathrm{PF:A})}.
\end{align}
$\sigma_{-1N}$ ($N=p$ or $n$) is given by
\beq
\sigma_{-1N}=\sigma_{-1N}^{(\mathrm{PF:A})}+\sigma_{-1N}^{(\mathrm{PF:B})}.
\eeq

\section{Results and discussion}
\label{sec3}

\subsection{Numerical inputs}
\label{sec31}

I employ the Dirac phenomenology~\cite{Ham90} to calculate
the optical potentials for the LP; the so-called
EDAD1 parameter set is used. The s.p. wave functions
for the nucleon inside A is calculated with the
Woods-Saxon potential by Bohr and Mottelson~\cite{BM69}. The nonlocal
correction proposed by Perey and Buck~\cite{PB62} is included
on both the distorted waves of the LP and the s.p.
wave functions of the target nucleon. The range of
nonlocality is taken to be 0.85~fm.

Although it is not explicitly shown in the formulae in
Sec.~\ref{sec2}, an $R$-dependent nucleon effective
mass $m_{N_{\mathrm{t}}}^*$, its so-called $k$-mass component, is adopted
in the numerical calculation as in Ref.~\cite{Oga03}.
The geometry of the difference between $m_{N_{\mathrm{t}}}^*$
and the bare mass $m_{N_{\mathrm{t}}}$ is assumed to be the same as that of
the Bohr-Mottelson potential. The ratio
$m_{N_{\mathrm{t}}}^*/m_{N_{\mathrm{t}}}$
at the center of the nucleus is taken to be 0.7 (0.8)
for proton (neutron), referring Mahaux and Sartor~\cite{MS91}.

As for $\tilde{t}$,
I adopt the $t$-matrix interactions of Franey and Love~\cite{FL85},
with the so-called final-energy prescription in making
an on-shell approximation. $\tilde{t}$ is, as defined,
evaluated in the nucleon-nucleon c.m. frame, and multiplied by
the M{\o }ller factor to be transformed to the nucleon-nucleus
c.m. frame. The neutron-neutron cross section is assumed to
be the same as the proton-proton one.

\subsection{Excitation energy distribution}
\label{sec32}

\begin{figure}[htbp]
\begin{center}
\includegraphics[width=0.4\textwidth,clip]{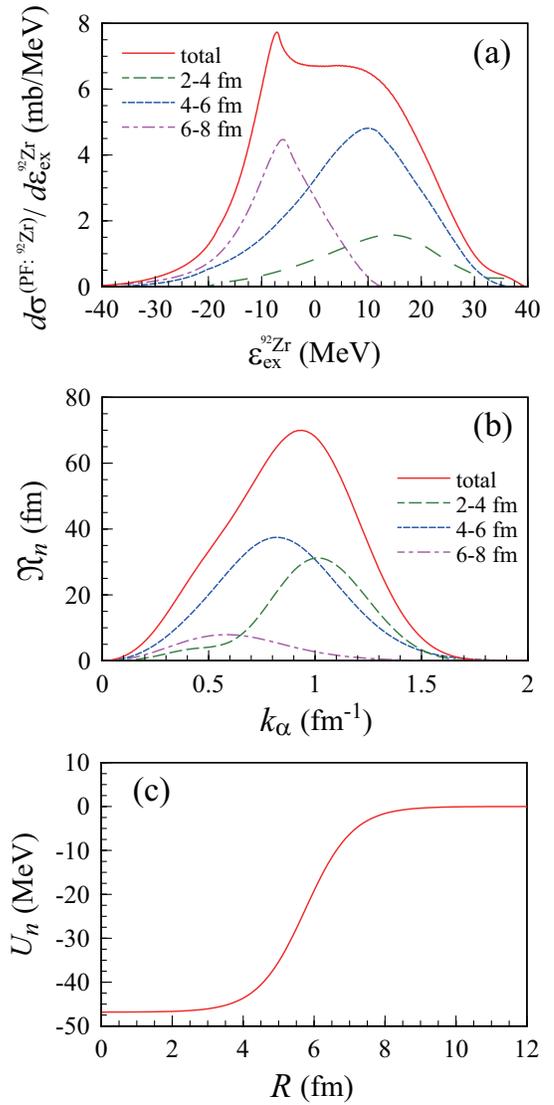}
\caption{(Color online)
(a) Excitation energy distribution of $^{92}$Zr after
$^{93}$Zr($p,p'x$) at 100~MeV (solid line).
The dashed, dotted, and dash-dotted lines show the contributions
from the $R=2$--4, 4--6, and 6--8~fm regions, respectively.
(b) Neutron momentum distribution in $^{93}$Zr (solid line);
the meaning of the other three lines is the same as in (a).
(c) Neutron single-particle potential.
}
\label{fig1}
\end{center}
\end{figure}
%
In Fig.~\ref{fig1}(a) I show by the solid line
$d\sigma^{(\mathrm{PF:}^{92}\mathrm{Zr})}
/d\ve_{\mathrm{ex}}^{{}^{92}\mathrm{Zr}}$
for $^{93}$Zr($p,p'x$) at 100~MeV/nucleon, without considering
the condition on
$\ve_{\mathrm{ex}}^{{}^{92}\mathrm{Zr}}$.
In other words, the result corresponds to the limit of
$\ve_{0n} \to -\infty$. This limit is always taken when I show
excitation energy distributions below.
The contributions from $R=2$--4, 4--6, and 6--8~fm are shown
by the dashed, dotted, and dash-dotted lines, respectively.
In Fig.~\ref{fig1}(b) the momentum distribution
$\mathfrak{N}_n$, which can be obtained by integrating
the WT over $\boldsymbol{R}$, of neutron in A is shown
by the solid line. I use the normalization
\beq
\int dk_\alpha \; \mathfrak{N}_n(k_\alpha)={\cal N}_{\mathrm{A}},
\eeq
where ${\cal N}_{\mathrm{A}}$ is the neutron number of A;
${\cal N}_{\mathrm{A}}=53$ in this case.
The other three lines in
Fig.~\ref{fig1}(b) represent the contributions from the
three regions of $R$. The correspondence between
the lines and the $R$ regions is the same as in Fig.~\ref{fig1}(a).
The neutron s.p. potential $U_n$ is plotted in Fig.~\ref{fig1}(c).

The dashed line in Fig.~\ref{fig1}(b) has a peak at
$k_\alpha \sim 1.0$~fm$^{-1}$, whereas the dotted and dash-dotted
ones at about 0.8 and 0.6~fm$^{-1}$, respectively.
This is consistent with the picture of the local Fermi-gas
model, that is, the upper limit of the nucleon momentum is
restricted by the local density in which the nucleon exits.
It should be noted that in the INCL model~\cite{Bou02,Bou13}
the nucleon momentum distribution
in a nucleus is assumed to be almost opposite to the results
in Fig.~\ref{fig1}(b); in the nuclear surface
a nucleon has to have high momenta. A modification on this
assumption
was proposed in Ref.~\cite{Man15}, which allows the nucleon to have
a lower momentum. This is more consistent with the result obtained
with the WT of the OBDM. However, there remains a somewhat
large difference between the momentum distributions in
Ref.~\cite{Man15} and Fig.~\ref{fig1}(b).

Neutron at smaller $R$ thus has larger kinetic energies.
On the other hand, $U_n$ becomes deeper as $R$ decreases as shown
in Fig.~\ref{fig1}(c). For $R=$2--4, the kinetic energy
at the peak of $\mathfrak{N}_n$ is about 25~MeV, whereas
$U_n\sim -47$~MeV; note that
$m_{N_{\mathrm{t}}}^*/m_{N_{\mathrm{t}}} \sim 0.8$ in this region.
Consequently, through Eq.~(\ref{eexb}) with
$S_n^{{}^{93}\mathrm{Zr}}=-8.28$~MeV,
$d\sigma^{(\mathrm{PF:}^{92}\mathrm{Zr})}
/d\ve_{\mathrm{ex}}^{{}^{92}\mathrm{Zr}}$
has a peak at around 14~MeV. If we go larger $R$, both the kinetic
energy and the depth of $U_n$ becomes smaller. Because the
$R$-dependence of the latter is stronger, at large $R$,
$\ve_\alpha$ can be positive and even exceed $S_n^{{}^{93}\mathrm{Zr}}$.
As a result,
$d\sigma^{(\mathrm{PF:}^{92}\mathrm{Zr})}
/d\ve_{\mathrm{ex}}^{{}^{92}\mathrm{Zr}}$
has a peak at negative $\ve_{\mathrm{ex}}^{{}^{92}\mathrm{Zr}}$,
as shown by
the dash-dotted line in Fig.~\ref{fig1}(a). Furthermore,
even though the dashed and dotted lines in Fig.~\ref{fig1}(a)
have a peak at positive
$\ve_{\mathrm{ex}}^{{}^{92}\mathrm{Zr}}$, they
are broad enough to have non-negligible cross sections for negative
$\ve_{\mathrm{ex}}^{{}^{92}\mathrm{Zr}}$.

\begin{figure}[htbp]
\begin{center}
\includegraphics[width=0.4\textwidth,clip]{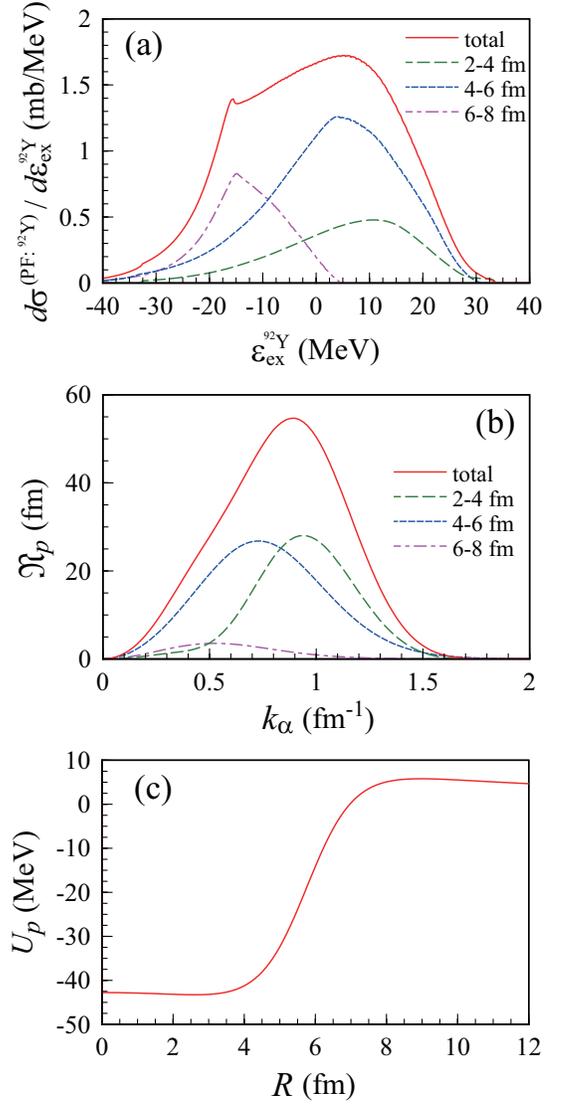}
\caption{(Color online)
Same as in Fig.~\ref{fig1} but for $^{92}$Y and for
proton in $^{93}$Zr.
}
\label{fig2}
\end{center}
\end{figure}
%
In Fig.~\ref{fig2} I show the results for the case in which
a target nucleon is proton, in the same way as in Fig.~\ref{fig1}.
The proton momentum distribution $\mathfrak{N}_p$ is similar to
$\mathfrak{N}_n$ in Fig.~\ref{fig1}(b); the difference in the magnitude
reflects that between the proton and neutron numbers of $^{93}$Zr.
The excitation energy distribution
$d\sigma^{(\mathrm{PF:}^{92}\mathrm{Y})}
/d\ve_{\mathrm{ex}}^{{}^{92}\mathrm{Y}}$
in Fig.~\ref{fig2}(a) is found to be slightly shifted to lower
energies, compared with
$d\sigma^{(\mathrm{PF:}^{92}\mathrm{Zr})}
/d\ve_{\mathrm{ex}}^{{}^{92}\mathrm{Zr}}$
in Fig.~\ref{fig1}(a). This is mainly due to the Coulomb interaction
in $U_p$ that increases $\ve_\alpha$.
One finds the absolute value of
$d\sigma^{(\mathrm{PF:}^{92}\mathrm{Y})}
/d\ve_{\mathrm{ex}}^{{}^{92}\mathrm{Y}}$
is somewhat smaller than that of
$d\sigma^{(\mathrm{PF:}^{92}\mathrm{Zr})}
/d\ve_{\mathrm{ex}}^{{}^{92}\mathrm{Zr}}$.
This comes from not only the difference between the proton and
neutron numbers of $^{93}$Zr but also that between the proton-proton
and proton-neutron total cross sections.

Interpretation of
$d\sigma^{(\mathrm{PF:B})}/d\ve_{\mathrm{ex}}^{\mathrm{B}}$,
where B is $^{92}$Zr or $^{92}$Y, for negative excitation energies
is not so trivial.
In this study, as mentioned in Sec.~\ref{sec2}, I introduce
$\ve_{0N_{\mathrm{t}}}$ as a lower limit of the integration, above which
$d\sigma^{(\mathrm{PF:B})}
/d\ve_{\mathrm{ex}}^{\mathrm{B}}$
contribute to
$\sigma_{-1N_{\mathrm{t}}}^{(\mathrm{PF:B})}$. I show in Table~\ref{tab1}
$\ve_{0n}$ and $\ve_{0p}$ determined so as to reproduce
the experimental values of
$\sigma_{-1n}^{(\mathrm{PF:B})}$ and $\sigma_{-1p}^{(\mathrm{PF:B})}$.
It should be noted that
$\sigma_{-1p}^{(\mathrm{PF:B})}$ is governed by
$\ve_{0p}$, whereas $\sigma_{-1n}^{(\mathrm{PF:B})}$ is by not only
$\ve_{0n}$ but also $\ve_{0p}$ through the neutron emission from A
after a $pp$ collision.
%
\begin{table}[htb]
\caption{
Threshold parameters $\ve_{0n}$ and $\ve_{0p}$ in unit of MeV.
The experimental data for one-nucleon knockout cross sections used
in the parameter fitting are taken from the references shown
in the table.
}
\begin{center}
\begin{tabular}{c|c|c|c} \hline\hline
                  & $\ve_{0n}$           & $\ve_{0p}$            & reference    \\ \hline
$^{90}$Sr@185MeV  & $-6.8_{-1.2}^{+1.3}$ & $ -3.2_{-1.8}^{+1.6}$ & \cite{Wan16} \\
$^{93}$Zr@100MeV  & $-3.0_{-0.9}^{+0.9}$ & $  3.3_{-1.3}^{+1.3}$ & \cite{Kaw17} \\
$^{107}$Pd@196MeV & $-9.2_{-1.4}^{+1.2}$ & $-10.9_{-1.2}^{+1.2}$ & \cite{Wan17} \\
$^{107}$Pd@118MeV & $-8.8_{-0.6}^{+0.6}$ & $ -2.6_{-1.7}^{+1.2}$ & \cite{Wan17} \\
$^{137}$Cs@185MeV & $-5.7_{-1.5}^{+1.4}$ & $  5.1_{-1.5}^{+2.2}$ & \cite{Wan16} \\ \hline\hline
\end{tabular}
\label{tab1}
\end{center}
\end{table}
In some cases
$\ve_{0N_{\mathrm{t}}}$ is found to be positive. This also will indicate
inadequacy of understanding of the properties of
nuclear many-body systems in the continuum.
Nevertheless, $\ve_{0N_{\mathrm{t}}}$ is found to lie in a reasonable
range, except $\ve_{0p}$ for $^{107}$Pd($p,p'x$) at 196~MeV.

\subsection{Energy dependence of proton-induced
one-nucleon knockout cross sections}
\label{sec33}

\begin{figure}[htbp]
\begin{center}
\includegraphics[width=0.4\textwidth,clip]{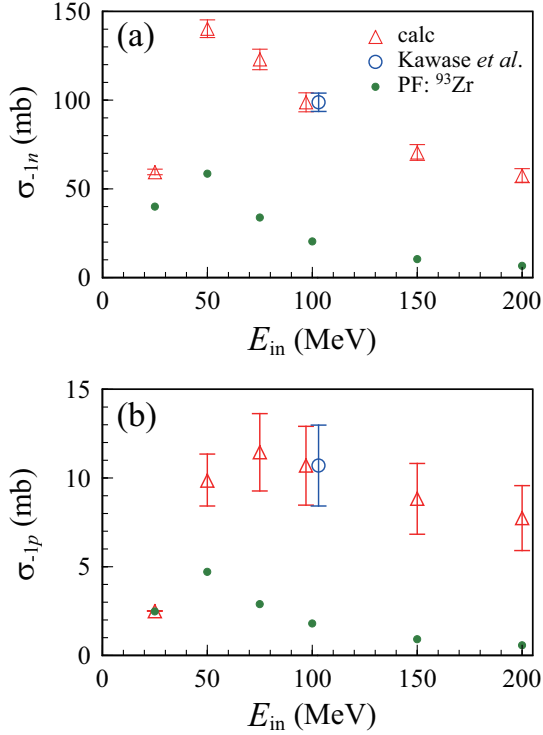}
\caption{(Color online)
One-neutron ($\sigma_{-1n}$) and one-proton ($\sigma_{-1p}$)
inclusive knockout cross sections (triangles) for
$^{93}$Zr($p,p'x$), as a function of the
incident energy $E_\mathrm{in}$.
The dots represent the contributions of the process in which
the pre-fragment (PF) is $^{93}$Zr.
The experimental data (circles) are taken from Ref.~\cite{Kaw17}.
}
\label{fig3}
\end{center}
\end{figure}
%
I show in Fig.~\ref{fig3} the dependence of (a) $\sigma_{-1n}$ and
(b) $\sigma_{-1p}$ on the incident energy $E_\mathrm{in}$
for $^{93}$Zr($p,p'x$).
The circles are the experimental data~\cite{Kaw17} and the
triangles are calculated results. The dots represent the
contributions of
$\sigma_{-1N}^{(\mathrm{PF:}^{93}\mathrm{Zr})}$ corresponding to
the mean values of $\sigma_{-1N}$.
I put a width of
$\ve_{0p}$ and $\ve_{0n}$ corresponding to the uncertainties
of the experimental data, and thereby estimate theoretical
uncertainties. For visibility in the plot I slightly shift the
energy of the results at 100~MeV; a similar shift is done
also in the following when needed.
One sees from Fig.~\ref{fig3}
that for $E_\mathrm{in} \ge 75$~MeV $\sigma_{-1n}$ and $\sigma_{-1p}$
decrease gradually as $E_\mathrm{in}$ increases, whereas 
they drop rather steeply at lower incident energy.

This energy dependence can qualitatively be understood as follows.
Let us describe the proton-induced one-nucleon
knockout process for a nucleus A with a $p+\mathrm{B}+N_\mathrm{t}$
three-body model. We first make the adiabatic approximation to the
$\mathrm{B}+N_\mathrm{t}$ motion. Then we assume that the elastic
breakup process, in which B is in the ground state after breakup,
can be neglected. In this case, as one sees from Eqs.~(17) and (18)
of Ref.~\cite{Yah11},
\beq
\sigma_{-1N_\mathrm{t}} \approx \sigma_\mathrm{R}(\mathrm{A})
                              - \sigma_\mathrm{R}(\mathrm{B}),
\eeq
where $\sigma_\mathrm{R}(\mathrm{C})$ is the proton total
reaction cross section for the nucleus C. Thus, the energy
dependence of $\sigma_{-1N_\mathrm{t}}$
is essentially determined by that for the $p$-$N_\mathrm{t}$ cross section
$\sigma_{pN_\mathrm{t}}$.
At low energies, as $\sigma_\mathrm{R}(\mathrm{C})$ does,
$\sigma_{-1N_\mathrm{t}}$ drops steeply because of the Coulomb barrier.
For the description of the behavior at low energies, however, one needs
a more quantitative analysis.

\begin{figure}[htbp]
\begin{center}
\includegraphics[width=0.4\textwidth,clip]{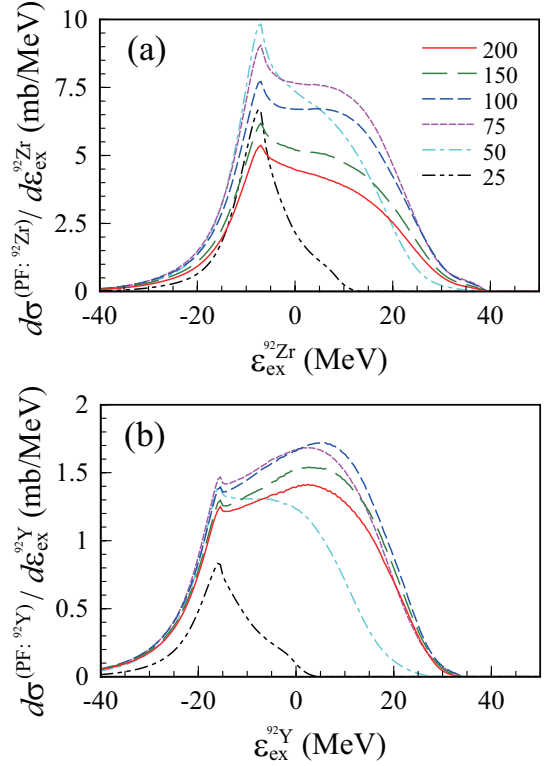}
\caption{(Color online)
(a) Excitation energy distribution of $^{92}$Zr after
$^{93}$Zr($p,p'x$) at 100~MeV. The solid, long-dashed, dashed,
dotted, dash-dotted, and dash-dot-dotted lines correspond to
$E_\mathrm{in}=200$, 150, 100, 75, 50, and 25~MeV, respectively.
(b) Same as (a) but for $^{92}$Y.
}
\label{fig4}
\end{center}
\end{figure}
%
Figure~\ref{fig4} shows
(a) $d\sigma^{(\mathrm{PF:}^{92}\mathrm{Zr})}
/d\ve_{\mathrm{ex}}^{{}^{92}\mathrm{Zr}}$
and (b)
$d\sigma^{(\mathrm{PF:}^{92}\mathrm{Y})}
/d\ve_{\mathrm{ex}}^{{}^{92}\mathrm{Y}}$.
In each panel, the solid, long-dashed, dashed, dotted,
dash-dotted, and dash-dot-dotted lines represent the results at
$E_\mathrm{in}=200$, 150, 75, 50, and 25~MeV, respectively.
For $E_\mathrm{in} \ga 75$~MeV, the shape of
the distributions does not change significantly, whereas the
absolute values monotonically decrease. The derivative for
the decrease with respect to $E_\mathrm{in}$ in panel (a)
is somewhat larger than that in panel (b). This is consistent
with the energy dependence of $\sigma_{pp}$ and $\sigma_{pn}$
in the energy region.

Below 75~MeV, the distributions in the positive excitation
energy region decrease rather steeply. At such low energies
the energy transfer is not large enough to knockout
a deeply bound nucleon that has small values of $\ve_\alpha$.
This restriction is more significant for proton because of
the Coulomb barrier.
The $E_\mathrm{in}$ dependence of the absolute values around
the peaks are determined by that of $\sigma_{pN_\mathrm{t}}$
and of the absorption due to the imaginary part of the
optical potential for proton. As a result of the competition
between these, for $^{92}$Zr ($^{92}$Y) the peak height decreases
$E_\mathrm{in}$ lower than 50~MeV (75~MeV).
It should be noted, however, that to understand the
energy dependence of $\sigma_{-1N}$ at low energies,
the contributions of
$\sigma_{-1N}^{(\mathrm{PF:}^{92}\mathrm{Zr})}$
also must be taken into consideration.

\begin{figure}[htbp]
\begin{center}
\includegraphics[width=0.4\textwidth,clip]{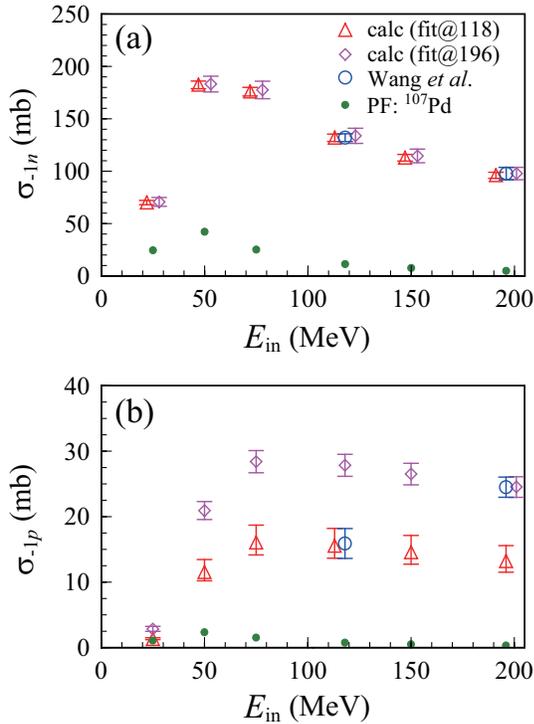}
\caption{(Color online)
Same as Fig.~\ref{fig3} but for $^{107}$Pd($p,p'x$).
The triangles (diamonds) represent the results calculated
with $\ve_{0n}$ and $\ve_{0p}$ determined
so as to reproduce the data at 118~MeV (196~MeV).
The experimental data are taken from Ref.~\cite{Wan17}.
}
\label{fig5}
\end{center}
\end{figure}
%
In Fig.~\ref{fig5} I show the $E_\mathrm{in}$ dependence of
(a) $\sigma_{-1n}$ and (b) $\sigma_{-1p}$ for $^{107}$Pd($p,p'x$).
The circles show the experimental data taken from Ref.~\cite{Wan17}.
The triangles (diamonds) represent the calculated results
with the threshold parameters ($\ve_{0n}$, $\ve_{0p}$) determined so as to
reproduce the experimental values at 118~MeV (196~MeV).
Features of the results are the same as in Fig.~\ref{fig3}.
An important remark on this system is that for $\sigma_{-1n}$
the two sets of theoretical results with different choices
of ($\ve_{0n}$, $\ve_{0p}$) agree well with each other.
This indicates that the $E_\mathrm{in}$ dependence of
$\sigma_{-1n}$ is well described by the present reaction model.
On the other hand, for $\sigma_{-1p}$ one finds a significant
difference between the two sets of results.
As mentioned above, the $E_\mathrm{in}$
dependence of $\sigma_{-1p}$ at high energies is expected to be
governed by that of $\sigma_{pp}$.
The experimental data show a completely different $E_\mathrm{in}$
dependence. It will be very interesting and important to understand
the origin of this behavior of the experimental data.

\subsection{Neutron-induced one-nucleon knockout cross sections}
\label{sec34}

\begin{figure}[htbp]
\begin{center}
\includegraphics[width=0.4\textwidth,clip]{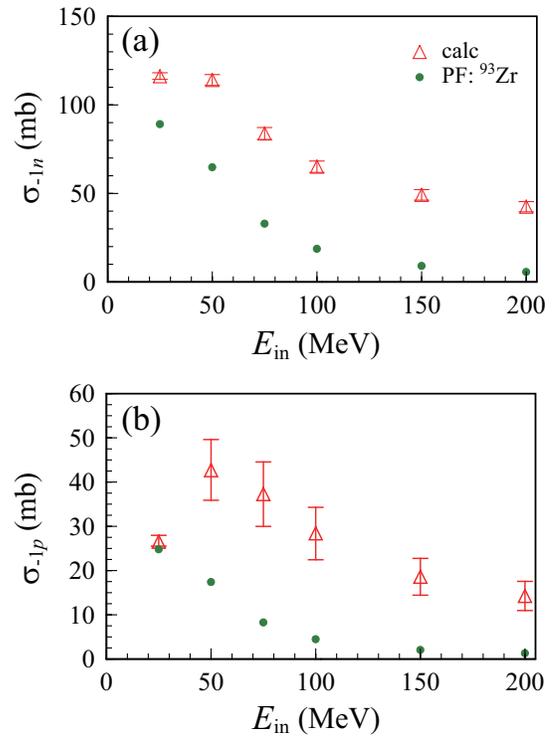}
\caption{(Color online)
Same as in Fig.~\ref{fig3} but for $^{93}$Zr($n,n'x$).
}
\label{fig6}
\end{center}
\end{figure}
%
\begin{figure}[htbp]
\begin{center}
\includegraphics[width=0.4\textwidth,clip]{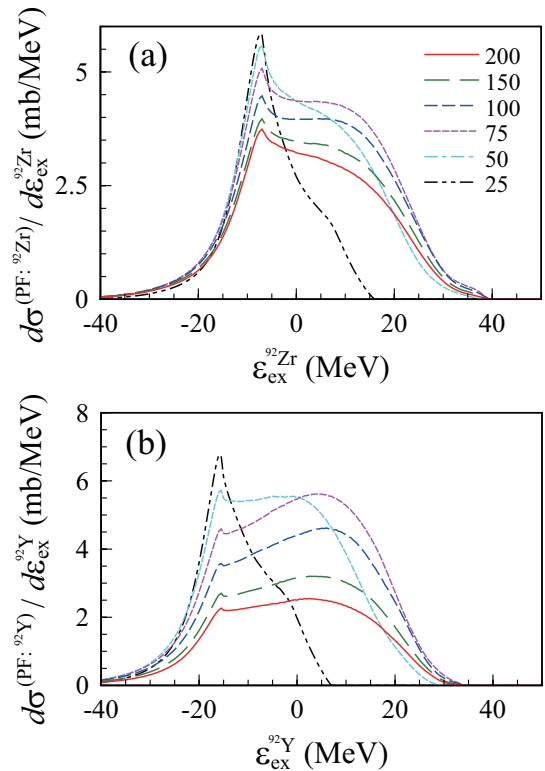}
\caption{(Color online)
Same as in Fig.~\ref{fig4} but for $^{93}$Zr($n,n'x$).
}
\label{fig7}
\end{center}
\end{figure}
%
Figures~\ref{fig6} and \ref{fig7} show the results for
$^{93}$Zr($n,n'x$), in the same way as in Figs.~\ref{fig3} and \ref{fig4},
respectively. In comparison with the results for $^{93}$Zr($p,p'x$)
one sees that the $E_\mathrm{in}$ dependence at low energies
becomes weaker. This is mainly due to the absence of the
Coulomb barrier for the LP. In fact, the peak height in
each panel of Fig.~\ref{fig7} increases as $E_\mathrm{in}$ decreases,
which indicates the energy dependence of
$\sigma_{nN_\mathrm{t}}$ is stronger than that of the absorption
caused by the neutron optical potential down to 25~MeV.
Another finding is the difference in the $E_\mathrm{in}$ dependence
at higher energy.
$\sigma_{-1p}$ depends on $E_\mathrm{in}$ more strongly
than $\sigma_{-1n}$, which is opposite to the behavior for
$(p,p'x)$.
This can easily be understood by the difference in the roles
of $\sigma_{nn}$ and $\sigma_{pn}$;
for $(n,n'x)$ the former (latter) that has a weaker (stronger)
$E_\mathrm{in}$ dependence contributes to
$\sigma_{-1n}$ ($\sigma_{-1p}$).
This explains also the difference in the ratios
$\sigma_{-1n}/\sigma_{-1p}$ for $(p,p'x)$ and $(n,n'x)$
except at low energies.

\begin{figure}[htbp]
\begin{center}
\includegraphics[width=0.4\textwidth,clip]{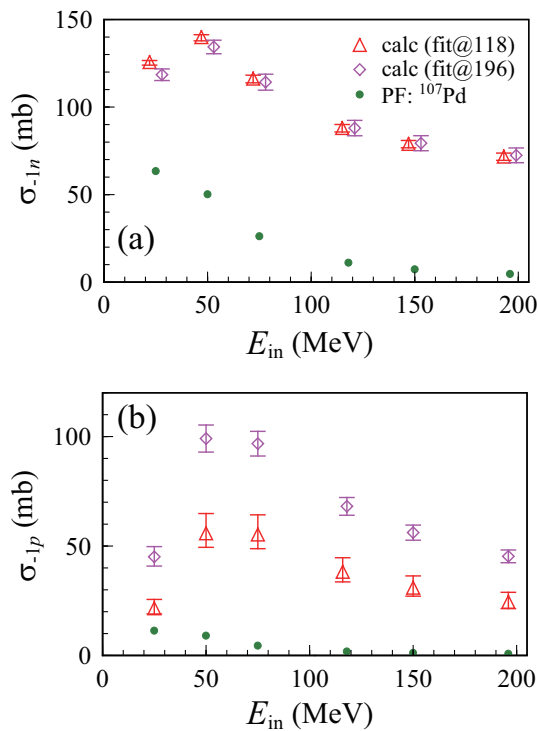}
\caption{(Color online)
Same as in Fig.~\ref{fig5} but for $^{107}$Pd($n,n'x$).
}
\label{fig8}
\end{center}
\end{figure}
%
Figure~\ref{fig8} shows the results for $^{107}$Pd($n,n'x$).
Features of the results in comparison with those in Fis.~\ref{fig5}
can be explained as in the same way as above.
The calculated $\sigma_{-1p}$ have large uncertainty,
reflecting the difference between the two sets of ($\ve_{0n}$, $\ve_{0p}$).

\section{Summary}
\label{sec4}

I have proposed a reaction model for describing nucleon-induced
inclusive one-nucleon knockout reactions. As an advantage to
the existing INCL models, the model incorporates the WT of OBDM of
a nucleus as a radial and momentum distribution of the struck
nucleon inside the nucleus. The proposed model contains threshold
parameters for the excitation energy distributions,
$\ve_{0n}$ and $\ve_{0p}$,
which are determined so that the one-neutron ($\sigma_{-1n}$)
and one-proton ($\sigma_{-1p}$) knockout cross sections
reproduce the experimental data. These parameters are considered
to reflect inadequacy of understanding of the nuclear many-body
wave functions, those in the continuum states in particular.

After fixing the threshold parameters, the model has been applied to the
study of the dependence of $\sigma_{-1n}$ and $\sigma_{-1p}$ on
the incident energy $E_\mathrm{in}$.
At energies higher than about 75~MeV, the energy dependence
is governed by that of the nucleon-nucleon total cross sections.
This picture is supported by the experimental data for
$\sigma_{-1n}$ of $^{107}$Pd($p,p'x$) measured at 118 and 196~MeV,
but contradicts the behavior of $\sigma_{-1p}$ of the reaction.
At low energies, because of the limitation of the energy transfer,
it becomes difficult to knockout a target nucleon in general.
However, the $E_\mathrm{in}$ dependence is determined by also
that of the nuclear absorption and the Coulomb penetrability
for the LP, and that of the nucleon-nucleon total cross sections.
The $E_\mathrm{in}$ dependence of $\sigma_{-1n}$ and $\sigma_{-1p}$
for ($n,n'x$) turned out to be quite different from that for
($p,p'x$). This can be crucial for {\lq\lq}extrapolating''
proton-induced spallation cross sections to
neutron-induced data that are inevitable for nuclear transmutation
studies.

In this study I used a phenomenological s.p. wave functions
of nuclei. Incorporation of more sophisticated nuclear
wave functions calculated by, e.g., density functional theory
will be desired. Combining a statistical model to describe
decay of reaction residues is another important future work.

\section*{Acknowledgments}

The author thanks
M.~Aikawa, S.~Ebata, N.~Furutachi, O.~Iwamoto, S.~Kawase, F.~Minato,
K.~Minomo, T.~Nakatsukasa, K.~Niita, H.~Otsu, H.~Sakurai, S.~Simoura,
H.~Wang, K.~Washiyama, and Y.~Watanabe
for valuable discussions and comments.
The computation was carried out with the computer facilities
at the Research Center for Nuclear Physics, Osaka University.
This work was supported in part
by Grant-in-Aid of the Japan Society for the
Promotion of Science (Grant No. JP16K05352)
and by the ImPACT Program of the Council for Science, Technology and
Innovation (Cabinet Office, Government of Japan).



\begin{thebibliography}{00}

\bibitem{IAEA04}
IAEA Technical Reports Series No. {\bf 435} (2004).

\bibitem{Wan16}
H. Wang {\it et al.}, Phys. Lett. B {\bf 754}, 104 (2016).

\bibitem{Wan17}
H. Wang {\it et al.}, Prog. Theor. Exp. Phys. {\bf 2017}, 021D01 (2017).

\bibitem{Kaw17}
S. Kawase {\it et al.}, Prog. Theor. Exp. Phys. {\bf 2017}, 093D03 (2017).

\bibitem{PHITS}
T. Sato, K. Niita, N. Matsuda, S. Hashimoto, Y. Iwamoto, S. Noda, T. Ogawa, H.
Iwase, H. Nakashima, T. Fukahori, K. Okumura, T. Kai, S. Chiba, T. Furuta, L.
Sihver, Particle and Heavy Ion Transport Code System PHITS, version 2.52, J.
Nucl. Sci. Technol. 2013 Sep; 50: 913-923.

\bibitem{Man15}
D. Mancusi, A. Boudard, J. Carbonell, J. Cugnon, J.-C. David,and S. Leray,
Phys. Rev. C {\bf 91}, 034602 (2015).

\bibitem{Bou02}
A. Boudard, J. Cugnon, S. Leray, and C. Volant,
Phys. Rev. C {\bf 66}, 044615 (2002).

\bibitem{Bou13}
A. Boudard, J. Cugnon, J.-C. David, S. Leray, and D. Mancusi,
Phys. Rev. C {\bf 87}, 014606 (2013).

\bibitem{Par17}
C. Paradela {\it et al.},
Phys. Rev. C {\bf 95}, 044606 (2017).

\bibitem {LK91}
Y. L. Luo and M. Kawai,
Phys. Rev. C {\bf 43}, 2367 (1991).

\bibitem {KM92}
M. Kawai and H. A. Weidenm\"uller,
Phys. Rev. C {\bf 45}, 1856 (1992).

\bibitem {Wat99}
Y. Watanabe, R. Kuwata, Sun Weili, M. Higashi, H. Shinohara,
M. Kohno, K. Ogata, and M. Kawai,
Phys. Rev. C {\bf 59}, 2136 (1999).

\bibitem {Oga99}
K. Ogata, M. Kawai, Y. Watanabe, Sun Weili, and M. Kohno,
Phys. Rev. C {\bf 60}, 054605 (1999).

\bibitem{Sun99}
Sun Weili, Y.~Watanabe, M.~Kohno, K.~Ogata, and M.~Kawai,
Phys. Rev. C {\bf 60}, 064605 (1999).

\bibitem {Oga02}
K. Ogata, Y. Watanabe, Sun Weili, M. Kohno, and M. Kawai,
Nucl. Phys. A {\bf 703}, 152 (2002).

\bibitem {Oga03}
K. Ogata, Y. Watanabe, Sun Weili, M. Kohno, and M. Kawai,
Proceedings of the Kyudai-RCNP International Symposium on Nuclear
Many-Body and Medium Effects in Nuclear Interactions and
Reactions, Fukuoka, 2002, World Scientific, Singapore, 2003, p. 231.

\bibitem {CR77}
N. S. Chant and P. G. Roos,
Phys. Rev. C {\bf 15}, 57 (1977).

\bibitem {CR83}
N. S. Chant and P. G. Roos,
Phys. Rev. C {\bf 27}, 1060 (1983).

\bibitem {Wak17}
T. Wakasa, K. Ogata, and T. Noro,
Prog. Part. Nucl. Phys. {\bf 96}, 32 (2017).

\bibitem{Ser47}
R. Serber, Phys. Rev. {\bf 72}, 1114 (1947).

\bibitem{Nii95}
K. Niita, S. Chiba, T. Maruyama, T. Maruyama, H. Takada, T.
Fukahori, Y. Nakahara, and A. Iwamoto,
Phys. Rev. C {\bf 52}, 2620 (1995).

\bibitem{Kan12}
Y. Kanada-En'yo, M. Kimura, and A. Ono,
Prog. Theor. Exp. Phys. {\bf 2012}, 01A202 (2012).

\bibitem{Iwa15}
O. Iwamoto, N. Iwamoto, S. Kunieda, F. Minato, and K. Shibata,
Nuclear Data Sheets {\bf 131}, 259 (2016).

\bibitem{Ham90}
S.~Hama, B.~C.~Clark, E.~D.~Cooper, H.~S.~Sherif,
and R.~L.~Mercer, Phys. Rev. C {\bf 41} 2737 (1990) ; \\
E.~D.~Cooper, S.~Hama, B.~C.~Clark, and R.~L.~Mercer,
{\it ibid.} {\bf 47}, 297 (1993).

\bibitem{BM69}
A. Bohr and B. R. Mottelson,
{\it Nuclear Structure} (Benjamin, New York, 1969), Vol.~I.

\bibitem{PB62}
G.~Perey and B.~Buck,
Nucl. Phys. {\bf 32}, 353 (1962).

\bibitem{MS91}
C.~Mahaux and R.~Sartor,
in \textit{Advances in Nuclear Physics},
edited by J. W. Negele and E. Vogt(Plenum, New York, 1991) Vol. 20, p. 1.

\bibitem{FL85}
M.~A.~Franey and W.~G.~Love,
Phys. Rev. C{\bf 31}, 488 (1985).



\bibitem{Yah11}
M. Yahiro, K. Ogata, and K. Minomo,
Prog. Theor. Phys. {\bf 126}, 167 (2011).


\end{thebibliography}
\end{document}